\documentclass[a4paper,twocolumn,amsmath,amssymb,prb,preprintnumbers,superscriptaddress,showpacs,longbibliography]{revtex4-1}

\pdfoutput=1 

\usepackage{textcomp}
\usepackage{amsmath}
\usepackage{amssymb}
\usepackage{graphicx}

\usepackage{dcolumn}
\usepackage{bm}
\usepackage{epsfig}
\usepackage[usenames]{color}
\usepackage{xcolor}

\newif\ifcom
\newif\ifdel

\comtrue                    
\deltrue                      

\begin{document}
\title{Extracellular electrical stimulation by ferroelectric displacement current in the switching regime}

\author{M.~Becker}
\email{maximilian.becker@nmi.de}
\affiliation{%
NMI Natural and Medical Sciences Institute at the University of T\"ubingen,
Markwiesenstr.~55,
72770 Reutlingen, Germany}
\affiliation{%
Physikalisches Institut,  Center for Quantum Science (CQ) and LISA$^+$,
University of T\"ubingen,
Auf der Morgenstelle 14,
72076 T\"ubingen, Germany}

\date{\today}

\begin{abstract} 

We analyze the extracellular stimulation current and the charge injection capacitity (CIC) of microelectrodes coated with an insulating layer to prevent toxic electrochemical effects in bioelectronic applications.
We show for a microelectrode coated with an insulating ferroelectric layer, that the ferroelectric polarization current contributes to the extracellular stimulation current.
Depending on the remanent polarization $P_{\rm r}$ of the ferroelectric, the polarization current in the switching regime can increase the CIC by up to two orders of magnitude as compared to the commonly used extracellular capacitive stimulation with microelectrodes that are coated with a dielectric layer.

\end{abstract} 

\maketitle


Extracellular electrical stimulation of neurons and recording of neural activity are the basic principles for many \textit{in vivo} applications such as implantable neuroprosthetic devices \cite{Cogan08}.
Among them are cochlear implants, retinal implants, deep brain stimulation for the treatment of Parkinsons's disease and brain-machine interfaces \cite{Kral19,Zrenner11,Mathieson12,Yue20,Miocinovic13,Musk19}.
Moreover, extracellullar electrical stimulation and recording  are also key elements for \textit{in vitro} applications e.g. drug testing and the study of neural networks by microelectrode arrays (MEAs) \cite{Stett03,Frey09}.

Typically, electrically active implants and MEAs utilize conductive (metallic) microelelectrodes with diameter $<$ 100 $\mu$m to provide the electro--neural interface for extracellular recording and stimulation \cite{Cogan08}.
In general, the main advantage of conductive microelectrodes is their high charge injection capacity (CIC), which is defined as the amount of injected charge density across the electrode--elelctrolyte interface during the leading stimulation phase of an extracellular  current \cite{Cogan08}.
Exemplary materials for the fabrication of conductive microelectrodes are Pt for high-density MEA applications \cite{Frey09} and IrO$_2$ which is utilized e.g. in state-of-the-art retinal prostheses \cite{Mathieson12,Haas20} and which can provide a CIC typically in the range of 1 - 5 mC/cm$^2$ ~\cite{Cogan08,Maeng19,Haas20}.

The major disadvantage of conductive microelectrodes is the difficulty to avoid electrochemical reactions (Faradaic processes) at the electrode--electrolyte interface, which induce corrosion of electrodes and cell or tissue damage \cite{Merrill05} and hence affect detrimentally the safety and long-term-stability of implantable neuroprosthetic devices.
A possible solution to prevent toxic electrochemical effects is the utilization of microelectrodes coated with a (thin) dielectric material.
This approach was crucial for the first realization of a silicon--neuron junction, where $p$-doped areas of the utilized silicon chip were covered with a 10-nm-thick dielectric layer of SiO$_2$ to prevent Faradaic processes and to stimulate the attached neuron by purely capacitive currents across the dielectric--electrolyte interface \cite{Fromherz1995}.

Modern capacitive biochips for extracellular electrical stimulation are fabricated in complementary metal-oxide-semiconductor (CMOS) technology and include field-effect transistors to record electrical signals from individual neurons \cite{Fromherz1991}, which enables the bidirectional communication between neurons and active silicon chips \cite{Bertotti14}.
However, despite extensive research for dielectric coatings to enhance the purely capacitive stimulation with microelectrodes, their CIC is still in the range of $\sim$ 1 - 5 $\mu$C/cm$^2$ ~\cite{Wallrapp06,Schoen07,Eickenscheidt12,Bertotti14,Dollt20}, which is approximately three orders of magnitude lower than the CIC of generic conductive microelectrodes.
Since the stimulation-threshold for the injected charge density depends on the geometric area of the microelectrode \cite{Corna18}, the CIC of capacitive microelectrodes is too low to achieve efficient extracellular stimulation with small microelectrodes (diameter $\sim$ 30 $\mu$m) which requires a CIC in the range of 0.1 - 0.9 mC/cm$^2$ ~\cite{Corna18}, although a tight tissue--electrode contact can lower the stimulation-threshold \cite{Eickenscheidt12}.
As a consequence, implantable neuroprosthetic devices utilize conductive metal-based microelectrodes although the absence of toxic electrochemical effects is highly desired in active electrical implants.
In this letter, we introduce an approach to significantly increase the CIC of insulated microelectrodes for efficient and long-term stable bioelectronic interfacing of electrogenic cells or tissue.

In general, extracellular electrical stimulation is achieved by the flow of ionic current between a microelectrode in close proximity to the target excitable cell or tissue and a counter electrode immersed in the electrolyte, which contains the cell or tissue \cite{Cogan08}. 
The ionic current flow in proximity to the cell or tissue causes a depolarization of the cell membrane, which evokes an action potential above a specific threshold.
\begin{figure}[t] 
\includegraphics[width=1\columnwidth]{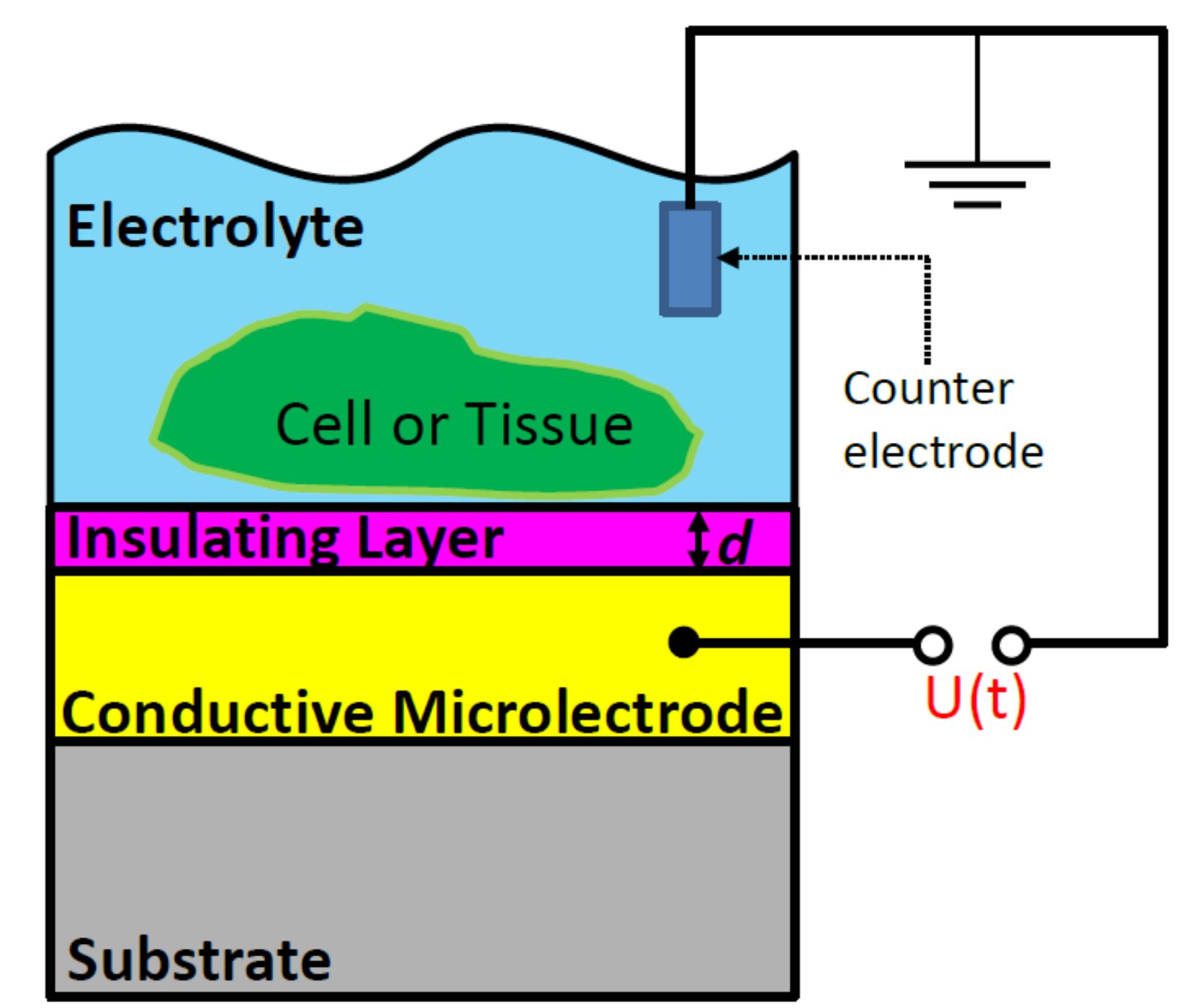}
\caption{Schematic experimental set-up for ectracellular electrical stimulation of electrogenic cells or tissue with a conductive microelectrode covered with an insulating layer of thickness $d$ (parallel-plate configuration) to prevent Faradaic processes at the electrode--electrolyte interface.
}
\label{fig:1} 
\end{figure} 
A schematic experimental set-up for the extracellular electrical stimulation of electrogenic cells or tissue with insulated microelectrodes  in a parallel-plate configuration with insulator thickness $d$ is depicted in Fig.~\ref{fig:1}.
Here, an applied voltage signal $U(t)$ generates an electric field $\mathbf{E}(t)$ across the insulating layer, which results in the occurrence of bound polarization charges at the surface of the insulator.
As a consequence, a flow of free charges in the electrolyte is generated to screen the bound polarization charges, and this ionic current corresponds to the extracellular stimulation current.
A simple stimulation signal is a voltage step $U(t)=U_0\Theta (t)$, where $\Theta$ denotes the Heaviside function.
Other generic voltage signals for extracellular stimulation are rectangular voltage pulses with stimulation time $T_{\rm stim}$ $\sim$ 1 ms, which are utilized in retinal implants \cite{Zrenner11} or deep brain stimulation \cite{Miocinovic13}.

In the following, we will analyze the experimental set-up for extracellular electrical stimulation with insulated microelectrodes [cf.  Fig.~\ref{fig:1}] within the framework of macroscopic electrodynamics \cite{Brandt_Book05}.
We make the assumption, that every vector quantity $\mathbf{V}$ is fully described by its component $V$ normal to the plane of the insulating layer and all other components vanish.
The focus of our analysis will be the insulating layer depicted in Fig.~\ref{fig:1}, which is described by the electric displacement field $\mathbf{D}$ defined as \cite{Brandt_Book05}
\begin{eqnarray} 
\mathbf{D}&=&\varepsilon_{0}\mathbf{E}+\mathbf{P} \quad,
\label{eq:D}
\end{eqnarray} 
with the vacuum permittivity $\varepsilon_{0}$ and electrical polarization $\mathbf{P}$.
We assume that the insulating layer has an infinite Ohmic resistance, which requires the absence of free charges, i.e. $\boldsymbol{\nabla}\cdot\mathbf{D}=0$.
We now consider Amp\`{e}re's law, which is part of the macroscopic Maxwell equations and is given by \cite{Brandt_Book05}
\begin{eqnarray} 
\boldsymbol{\nabla}\times\mathbf{H}&=&\mathbf{J}+\frac{\partial\mathbf{D}}{\partial t} \quad,
\label{eq:Ampere}
\end{eqnarray} 
with the magnetic field strength $\mathbf{H}$, the conduction current density $\mathbf{J}$ and the time derivative  $\partial\mathbf{D}/\partial t$, which corresponds to the density of Maxwell's displacement current.
Due to the insulation of the conductive microelectrode [cf.  Fig.~\ref{fig:1}] electrochemical (Faradaic) charge transport is suppressed, which results in a vanishing conduction current density $\mathbf{J}$ in Eq.~(\ref{eq:Ampere}).
Thus, the extracellular stimulation current density $\mathbf{J}_{\rm stim}$ provided by an insulated microelectrode corresponds to the displacement current density, i.e. $\mathbf{J}_{\rm stim }=\partial\mathbf{D}/\partial t$ and by definition, the CIC can be calculated according to
\begin{eqnarray} 
\rm{CIC}& = & \int_{0}^{T_{\rm stim}}\vert J_{\rm stim} \vert dt   \quad.
\label{eq:CIC}
\end{eqnarray} 

The common extracellular capacitive stimulation utilizes an insulating layer which consists of a material with solely dielectric properties.
In linear approximation, the electrical polarization of a dielectric material is given by \cite{Brandt_Book05}
\begin{eqnarray} 
\mathbf{P}(\mathbf{E})&=&\varepsilon_{0}(\varepsilon_{r}-1)\mathbf{E} \quad,
\label{eq:P_DE}
\end{eqnarray} 
where $\varepsilon_{r}$ denotes the relative permittivity of the material.
Inserting Eq.~(\ref{eq:P_DE}) in Eq.~(\ref{eq:D}) yields the linear relationship
\begin{eqnarray} 
\mathbf{D}(\mathbf{E})&=&\varepsilon_{0} \varepsilon_{r} \mathbf{E} \quad.
\label{eq:D_DE}
\end{eqnarray} 
By introducing the specific capacitance $C_s \equiv \varepsilon_{0}\varepsilon_{r}/d$, the resulting capacitive stimulation current density  in repsonse to a time-varying electric field $E(t)=U(t)/d$ is then given by
\begin{eqnarray} 
J_{\rm stim}&=&C_s \frac{\partial}{\partial t}U(t)  \quad .
\label{eq:j_DE}
\end{eqnarray} 
Thus, previous attempts to enhance the CIC of capacitive microelectrodes were based on the approach to increase the specific capacitance of the dielectric layer.

We now consider a material with ferroelectric properties.
Ferroelectrics are characterized by the existence of a (spontaneous) ferroelectric polarization $\mathbf{P}^{{\rm (FE)}}$, which can be reoriented by an applied electric field, i.e. $\mathbf{P}^{{\rm (FE)}}=\mathbf{P}^{{\rm (FE)}}(\mathbf{E})$  \cite{Waser_Book05}.
The finite value of the ferroelectric polarization at zero electric field is called the remanent polarization $P_{\rm r}$.
By applying an electric field above the coervice field $E_{\rm C}$ of the ferroelectric, the remanent polarization can be switched between the bistable states $\pm P_{\rm r}$, which is the basis for nonvolatile ferroelectric memory applications \cite{Park18}.
Here, the discovery of ferroelectricity in HfO$_2$-based thin films represents a real breakthrough due to the full CMOS compatibility of this material class and outstanding ferroelectric properties at the nanoscale \cite{Mueller11,Park18,Wei18}.
Moreover, conventional dielectric HfO$_2$ layers have been previously utilized in capacitive biochips \cite{Wallrapp06,Schoen07,Dollt20}, which demonstrates their biocompatibility.
Note, that materials with ferroelectric properties also exhibit dielectric properties but not vice versa.
Thus, the electrical polarization of a material with ferroelectric properties is given by $\mathbf{P}=\mathbf{P}^{{\rm (DE)}}+\mathbf{P}^{{\rm (FE)}}$.
As a consequence, it follows from Eq.~(\ref{eq:D})
\begin{eqnarray} 
\mathbf{D}(\mathbf{E})&=&\varepsilon_{0}\varepsilon_{r} \mathbf{E}+\mathbf{P}^{{\rm (FE)}}(\mathbf{E})  \quad
\label{eq:D_FE}
\end{eqnarray} 
and the resulting ferroelectric displacement current density for extracellular stimulation $J_{\rm stim}$ in repsonse to a time-varying electric field $E(t)=U(t)/d$ is then given by
\begin{eqnarray} 
J_{\rm stim}&=&C_s \frac{\partial}{\partial t}U(t)+\frac{\partial}{\partial t} P^{{\rm (FE)}}(U(t))  \quad,
\label{eq:J_FE}
\end{eqnarray} 
which exhibits the ferroelectric polarization current $\partial P^{{\rm (FE)}}/\partial t $ as an additional contribution as compared to the purely capacitive stimulation current density  Eq.~(\ref{eq:j_DE}).

Depending on the applied electric field strength $E(t)=U(t)/d$, we can identify two different regimes of the ferroelectric polarization current which are separated by the coercive field $E_{\rm C}$ of the ferroelectric, i.e.
\begin{eqnarray} 
\frac{\partial}{\partial t}P^{{\rm (FE)}}(E(t))&\equiv&\begin{cases}
J^{({\rm SSW)}}(t) & E<E_{{\rm C}}\\
J^{({\rm SW)}}(t) & E>E_{{\rm C}}
\end{cases}  \quad,
\label{eq:J_Regimes}
\end{eqnarray} 
where $J^{({\rm SSW)}}(t)$ corresponds to the subswitching regime and $J^{({\rm SW)}}(t)$ to the switching regime of the ferroelectric.
For subswitching fields above a specific threshold field $E_{\rm T}$ the linear approximation Eq.~(\ref{eq:P_DE}) is no longer valid for a ferroelectric material due to the irreversible motion of ferroelectric domain walls, which results in a nonlinear $P$ vs $E$ relationship described by the Rayleigh law \cite{Damjanovic1998}.
As a consequence, the ferroelectric displacement current density  Eq.~(\ref{eq:J_FE}) in response to a harmonic ac electric field can contain higher harmonics  due to the contribution from irreversible domain wall motion contained in $J^{({\rm SSW)}}(t)$ \cite{Miga07}, which might be interesting for extracellular stimulation with low-frequency sinusoidal voltage signals \cite{Freeman10}.
Experimentally, the subswitching regime of a ferroelectric can be analyzed in detail by impedance spectroscopy and equivalent-circuit fitting with the recently introduced domain wall pinning element $Z_{\rm DW}$ \cite{BeckerJAP20}.

In the following, we will focus our discussion on the ferroelectric switching regime $E > E_{\rm C}$. 
The corresponding switching kinetics is a very complex problem and there exist several models to explain the time-dependence of ferroelectric polarization reversal \cite{Jo07}.
The traditional approach to explain the switching kinetics is the Kolmogorov-Avrami-Ishibashi (KAI)-model \cite{Kolmogorov1937,Avrami1939,Ishibashi1971}, which is based on the classical statistical theory of nucleation and predicts for the time-dependent change in ferroelectric polarization $\Delta P^{{\rm (FE)}}(t)$ during switching between the bistable states $\pm P_{\rm r}$ , the relationship \cite{Jo07}
\begin{eqnarray} 
 \Delta P^{{\rm (FE)}}(t)&=&2P_{\rm r}\left\{1-\exp[-(t/t_0)^n]\right\} \quad,
\label{eq:P_KAI}
\end{eqnarray} 
with the effective dimension $n$ and the characteristic switching time $t_0$ as free parameters.
For thin films, the effective dimension is $n$ = 2 \cite{Jo07} and the time-dependence of the magnitude of the corresponding switching current density $J^{\rm (SW)}(t)$ is then given by 
\begin{eqnarray} 
\vert J^{{\rm (SW)}}(t) \vert&=&4P_{\rm r} (t/t_0^2) \exp[-(t/t_0)^2] \quad.
\label{eq:J_KAI}
\end{eqnarray} 
\begin{figure}[t] 
\includegraphics[width=1\columnwidth]{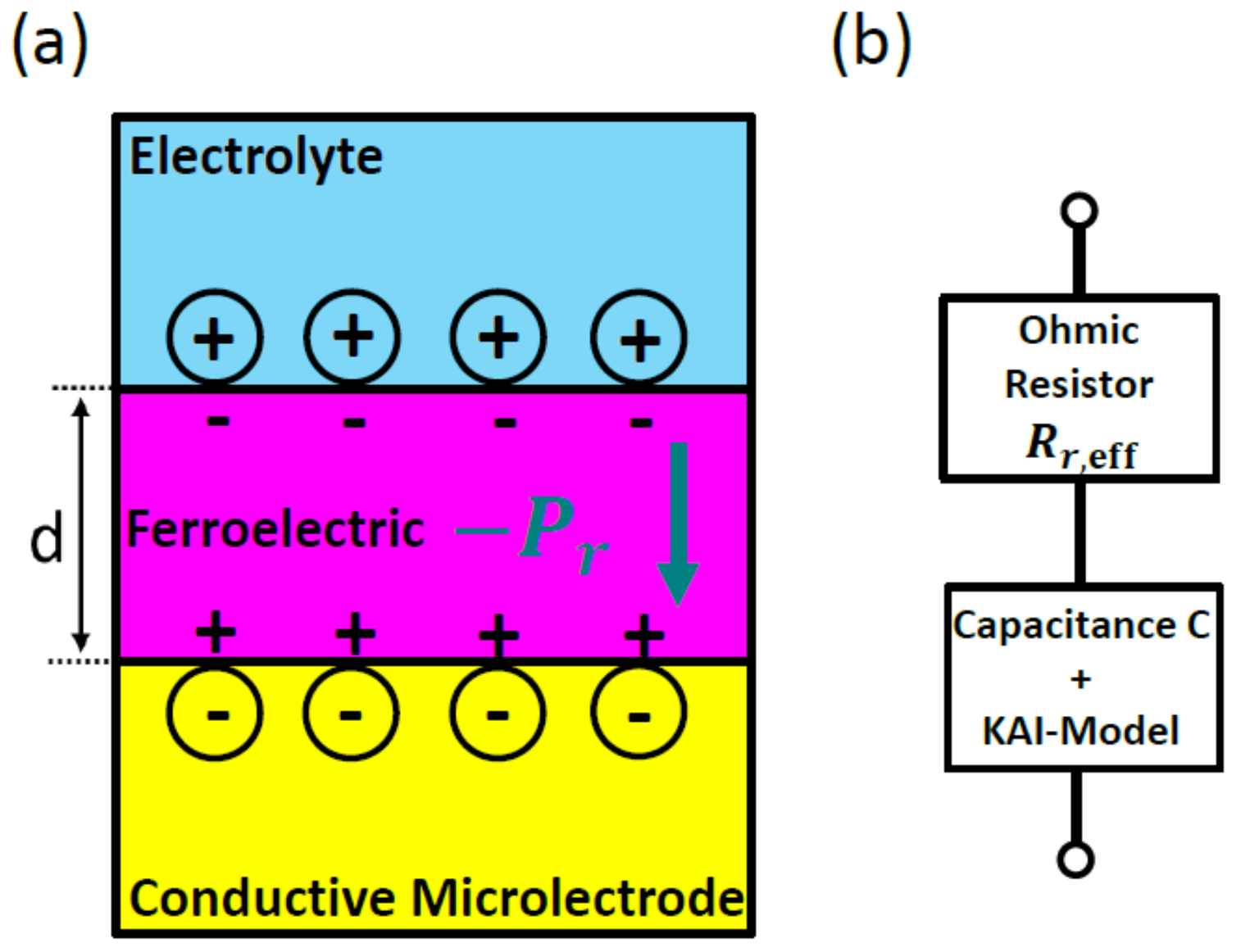}
\caption{(a) Electrolyte--ferroelectric--conductor (EFC) configuration in the polarization state -$P_{\rm r}$. In equilibrium, free charges in the electrolyte and in the conductive microelectrode screen the bound polarization charges at the surface of the ferroelectric. (b) Equivalent-circuit model of the EFC configuration which consists of an Ohmic resistor to model the resistance of the electrolyte and the electrodes and a capacitance in combination with the KAI-model which represents the ferroelectric layer.
}
\label{fig:2} 
\end{figure} 
\begin{table*}[t] 
\caption{Specifications to simulate the extracellular stimulation current response Eq.~(\ref{eq:J_stim}) of an insulated microelectrode (50 $\mu$m in diameter) in EFC configuration  [cf. Fig.~\ref{fig:2}] in the switching regime. The listed ferroelectric material properties correspond to a 9.5-nm-thick  Hf$_{0.5}$Zr$_{0.5}$O$_2$ layer \cite{Mueller11}. The resistivity $\rho$ of the electrolyte corresponds to the value of phosphate-buffered saline at room temperature \cite{Haas20}.}
\begin{ruledtabular}
\begin{tabular}{cccccccccc}

&$d$ (nm)  & $P_{\rm r}$ ($\mu$C/cm$^2$)   & $E_{\rm C}$ (MV/cm)   &$t_0$ ($\mu$s)    &$\varepsilon_{r}$  &r ($\mu$m) & $\rho$ ($\Omega$cm) &$U_0$ (V)     \\ \hline 

& 9.5          & 16	                                            & 1                                & 10             & 40	                              & 25	         &66.7                     &1
\end{tabular}
\end{ruledtabular}
\label{Tab-1}
\end{table*} 
We now reconsider the schematic experimental set-up for extracellular stimulation depicted in Fig.~\ref{fig:1} with the specification that the conductive microelectrode with radius $r$ and geometrical surface area $A=\pi r^2$ is insulated with a ferroelectric layer, which corresponds to the electrolyte--ferroelectric--conductor (EFC) configuration shown in Fig.~\ref{fig:2}(a).
To analyze the CIC and the time-dependence of the stimulation current density provided by the EFC configuration in the switching regime $J_{\rm stim}^{\rm EFC}(t)$, we describe the EFC configuration by the equivalent-circuit depicted  in Fig.~\ref{fig:2}(b). 
Here, the Ohmic resistance models the electrolyte and the electrodes and the capacitance $C=\varepsilon_{0}\varepsilon_{r}A/d$ in combination with the KAI-model models the ferroelectric layer of the EFC configuration.
The Ohmic resistance can be approximated by the effective resistance $R_{r, \rm{eff}}$ for a circular microelectrode with radius $r$ and an adjacent electrolyte with specific resistance $\rho$ according to \cite{Chen20}
\begin{eqnarray} 
R_{r, \rm{eff}}&=&\frac{\rho}{\pi r}\quad.
\label{eq:R_eff}
\end{eqnarray} 
The stimulation current density $J_{\rm stim}^{\rm EFC}(t)$ provided by the EFC configuration [cf.  Fig.~\ref{fig:2}]  in response to an applied voltage step with height $U_0$, which induces polarization reversal from the initial state -$P_{\rm r}$ to the final state +$P_{\rm r}$ is then given by
\begin{eqnarray} 
J_{\rm stim}^{\rm EFC}(t)&=&\frac{U_0}{r \rho}\exp[\frac{-t}{R_{r, \rm{eff}}C}] + 4P_{\rm r} (t/t_0^2) \exp[-(t/t_{0})^2] \quad, \nonumber \\
\label{eq:J_stim}
\end{eqnarray} 
where the first term in Eq.~(\ref{eq:J_stim}) is the capacitive transient $J^{(\rm C)}(t)$.
Note, that ferroelectric polarization reversal in EFC configuration has been demonstrated for various ferroelectric materials \cite{Ferris13, Fabiano14,Toss17}.
\begin{figure}[t] 
\includegraphics[width=1\columnwidth]{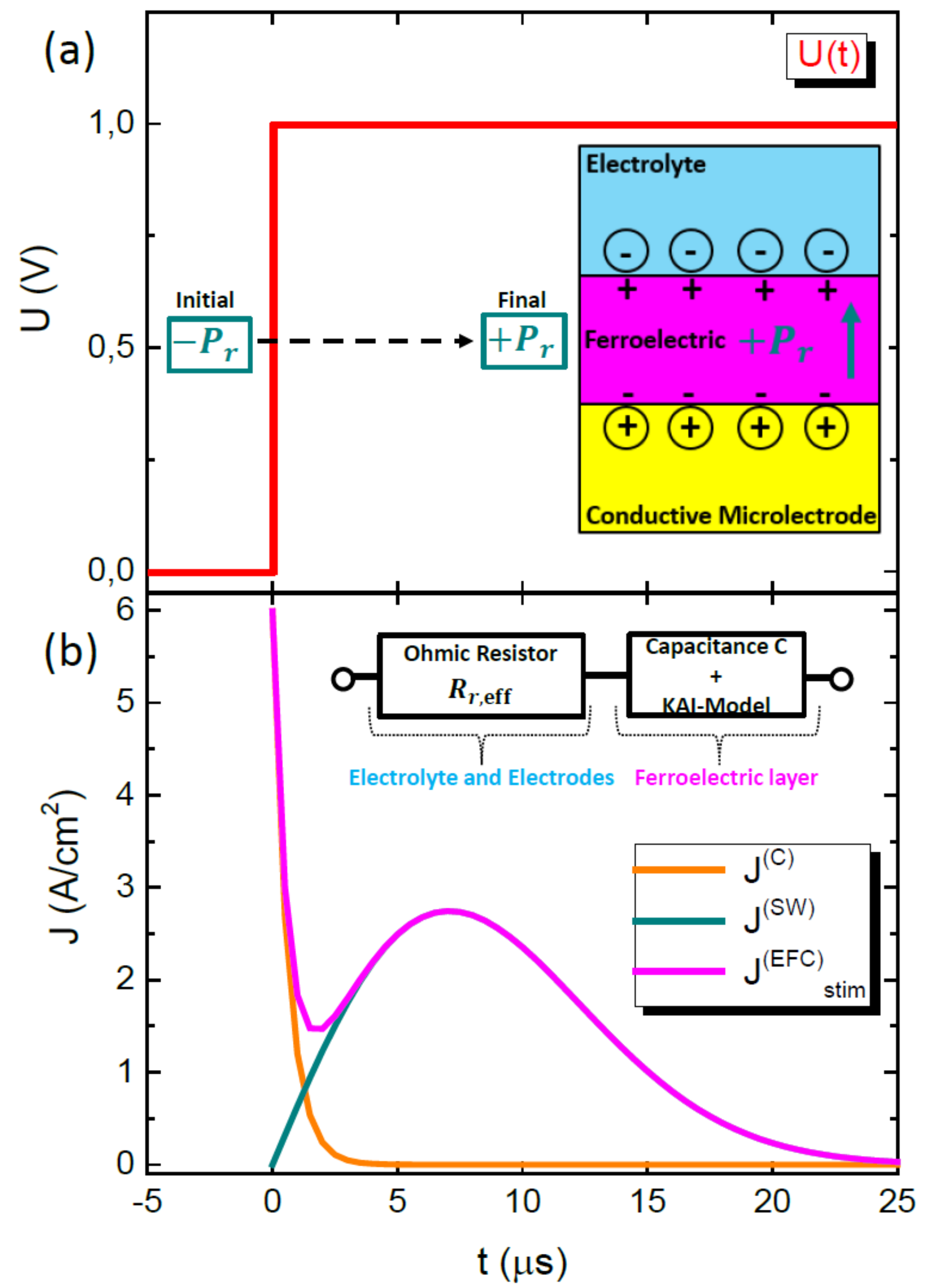}
\caption{(a) Voltage step with height $U_0$ = 1 V applied to the EFC configuration [cf. inset and  Fig.~\ref{fig:2}], to switch the remanent polarization from the initial state -$P_{\rm r}$ to the final state +$P_{\rm r}$.  (b) Simulated extracellular current response during polarization reversal of the EFC configuration according to Eq.~(\ref{eq:J_stim}) with the specifications summarized in Table \ref{Tab-1}.
}
\label{fig:3} 
\end{figure} 

To simulate the extracellular stimulation current response of an insulated microelectrode in EFC configuration [cf.  Fig.~\ref{fig:2}] during polarization reversal, we utilize Eq.~(\ref{eq:J_stim}) with the specifications summarized in Table \ref{Tab-1}.
The applied voltage step with height $U_0$ = 1 V depicted in Fig.~\ref{fig:3}(a), generates an electric field above the coercive field of the ferroelectric layer [cf. Table \ref{Tab-1}] and the resulting current response $J_{\rm stim}^{\rm EFC}(t)$ in the switching regime is shown in Fig.~\ref{fig:3}(b).
Here, the capacitive transient $J^{(\rm C)}(t)$ is superimposed by the bell-shaped switching current density $J^{{\rm (SW)}}(t)$ and the resulting CIC Eq.~(\ref{eq:CIC}) for complete switching during the stimulation time (i.e. $T_{\rm stim}>t_0$) is given by \cite{Damjanovic1998}
\begin{eqnarray} 
\rm{CIC}&=&C_sU_0 + 2P_{\rm r}  \quad,
\label{eq:CIC_FE}
\end{eqnarray} 
which exhibits the additional contribution $2P_{\rm r}$ compared to the CIC of microelectrodes with dielectric coating for extracellular capacitive stimulation \cite{Fromherz1995}.
Note that repetetive stimulation with a CIC according to  Eq.~(\ref{eq:CIC_FE}) requires a subsequent voltage pulse of opposite polarity to reorient the remanent polarization $P_{\rm r}$ back to its initial state.
This is not necessary for antiferroelectric materials or ferroelectrics in the paraelectric phase, where the remanent polarization is zero and which might be interesting alternatives for stimulation with solely unipolar voltage signals.

For a ferroelectric microelectrode with the specifications listed in Table \ref{Tab-1}, Eq.~(\ref{eq:CIC_FE}) yields a CIC of 35.7 $\mu$C/cm$^2$, which is dominated by the contribution $2P_{\rm r}$.
Therefore, another interesting CMOS compatible material class is Al$_{1-x}$Sc$_{x}$N, which exhibit ferroelectric properties with $P_{\rm r}$ up to 110 $\mu$C/cm$^2$ \cite{Fichtner19} resulting in a CIC above 200 $\mu$C/cm$^2$, which is in the range of previously reported stimulation-thresholds for small conductive microelectrodes (30 $\mu$m in diameter) \cite{Corna18} and which is approximately two orders of magnitude above the CIC of state-of-the-art capacitive microelectrodes \cite{Wallrapp06,Schoen07,Eickenscheidt12,Bertotti14,Dollt20}.
Note, that the CIC of ferroelectric microelectrodes might be further increased by the fabrication of 3D structures \cite{Park18}.

In conclusion, our results pave the way to utilize ferroelectrics for bioelectronic interfacing of electrogenic cells or tissue. 
The resulting ferroelectric--neural interface is a promising approach for extracellular electrical stimulation without toxic electrochemical effects, which is crucial for implantable neuroprosthetic devices such as retinal-implants or brain-machine interfaces.
Clearly, substantial future work is needed to develop stimulus protocols for specific neuronal classes and for different ferroelectric materials with different remanent polarization and switching kinetics.
Additional possibilities for future studies on ferroelectric--neural interfaces are extracellular electrical recordings by utilizing the ferroelectric layer as gate oxide of a ferroelectric field-effect transistor (FeFET), which polarization state is sensitive to extracellular potentials generated by adjacent neurons.
This might result in the coupling of artificial neurons based on FeFETs \cite{Mulaosmanovic18} with biological neurons for future neuromorphic computing applications.

The author thanks Claus Burkhardt and Dieter Koelle for very fruitful discussions.

\section*{Data availability}
The data that supports the findings of this study are available within the article.



\clearpage

\bibliography{FE-Stim}

\begin{thebibliography}{39}%
\makeatletter
\providecommand \@ifxundefined [1]{%
 \@ifx{#1\undefined}
}%
\providecommand \@ifnum [1]{%
 \ifnum #1\expandafter \@firstoftwo
 \else \expandafter \@secondoftwo
 \fi
}%
\providecommand \@ifx [1]{%
 \ifx #1\expandafter \@firstoftwo
 \else \expandafter \@secondoftwo
 \fi
}%
\providecommand \natexlab [1]{#1}%
\providecommand \enquote  [1]{``#1''}%
\providecommand \bibnamefont  [1]{#1}%
\providecommand \bibfnamefont [1]{#1}%
\providecommand \citenamefont [1]{#1}%
\providecommand \href@noop [0]{\@secondoftwo}%
\providecommand \href [0]{\begingroup \@sanitize@url \@href}%
\providecommand \@href[1]{\@@startlink{#1}\@@href}%
\providecommand \@@href[1]{\endgroup#1\@@endlink}%
\providecommand \@sanitize@url [0]{\catcode `\\12\catcode `\$12\catcode
  `\&12\catcode `\#12\catcode `\^12\catcode `\_12\catcode `\%12\relax}%
\providecommand \@@startlink[1]{}%
\providecommand \@@endlink[0]{}%
\providecommand \url  [0]{\begingroup\@sanitize@url \@url }%
\providecommand \@url [1]{\endgroup\@href {#1}{\urlprefix }}%
\providecommand \urlprefix  [0]{URL }%
\providecommand \Eprint [0]{\href }%
\providecommand \doibase [0]{http://dx.doi.org/}%
\providecommand \selectlanguage [0]{\@gobble}%
\providecommand \bibinfo  [0]{\@secondoftwo}%
\providecommand \bibfield  [0]{\@secondoftwo}%
\providecommand \translation [1]{[#1]}%
\providecommand \BibitemOpen [0]{}%
\providecommand \bibitemStop [0]{}%
\providecommand \bibitemNoStop [0]{.\EOS\space}%
\providecommand \EOS [0]{\spacefactor3000\relax}%
\providecommand \BibitemShut  [1]{\csname bibitem#1\endcsname}%
\let\auto@bib@innerbib\@empty
\bibitem [{\citenamefont {Cogan}(2008)}]{Cogan08}%
  \BibitemOpen
  \bibfield  {author} {\bibinfo {author} {\bibfnamefont {S.~F.}\ \bibnamefont
  {Cogan}},\ }\bibfield  {title} {\enquote {\bibinfo {title} {Neural
  stimulation and recording electrodes},}\ }\href@noop {} {\bibfield  {journal}
  {\bibinfo  {journal} {Annu. Rev. Biomed. Eng.}\ }\textbf {\bibinfo {volume}
  {10}},\ \bibinfo {pages} {275} (\bibinfo {year} {2008})}\BibitemShut
  {NoStop}%
\bibitem [{\citenamefont {Kral}\ \emph {et~al.}(2019)\citenamefont {Kral},
  \citenamefont {Dorman},\ and\ \citenamefont {Wilson}}]{Kral19}%
  \BibitemOpen
  \bibfield  {author} {\bibinfo {author} {\bibfnamefont {A.}~\bibnamefont
  {Kral}}, \bibinfo {author} {\bibfnamefont {M.~F.}\ \bibnamefont {Dorman}}, \
  and\ \bibinfo {author} {\bibfnamefont {B.~S.}\ \bibnamefont {Wilson}},\
  }\bibfield  {title} {\enquote {\bibinfo {title} {Neuronal development of
  hearing and language: Cochlear implants and critical periods},}\ }\href@noop
  {} {\bibfield  {journal} {\bibinfo  {journal} {Annu. Rev. Neurosci.}\
  }\textbf {\bibinfo {volume} {42}},\ \bibinfo {pages} {47} (\bibinfo {year}
  {2019})}\BibitemShut {NoStop}%
\bibitem [{\citenamefont {Zrenner}\ \emph {et~al.}(2011)\citenamefont
  {Zrenner}, \citenamefont {Bartz-Schmidt}, \citenamefont {Benav},
  \citenamefont {Besch}, \citenamefont {Bruckmann}, \citenamefont {Gabel},
  \citenamefont {Gekeler}, \citenamefont {Greppmaier}, \citenamefont
  {Harscher}, \citenamefont {Kibbel}, \citenamefont {Koch}, \citenamefont
  {Kusnyerik}, \citenamefont {Peters}, \citenamefont {Stingl}, \citenamefont
  {Sachs}, \citenamefont {Stett}, \citenamefont {Szurman}, \citenamefont
  {Wilhelm},\ and\ \citenamefont {Wilke}}]{Zrenner11}%
  \BibitemOpen
  \bibfield  {author} {\bibinfo {author} {\bibfnamefont {E.}~\bibnamefont
  {Zrenner}}, \bibinfo {author} {\bibfnamefont {K.~U.}\ \bibnamefont
  {Bartz-Schmidt}}, \bibinfo {author} {\bibfnamefont {H.}~\bibnamefont
  {Benav}}, \bibinfo {author} {\bibfnamefont {D.}~\bibnamefont {Besch}},
  \bibinfo {author} {\bibfnamefont {A.}~\bibnamefont {Bruckmann}}, \bibinfo
  {author} {\bibfnamefont {V-P.}\ \bibnamefont {Gabel}}, \bibinfo {author}
  {\bibfnamefont {F.}~\bibnamefont {Gekeler}}, \bibinfo {author} {\bibfnamefont
  {U.}~\bibnamefont {Greppmaier}}, \bibinfo {author} {\bibfnamefont
  {A.}~\bibnamefont {Harscher}}, \bibinfo {author} {\bibfnamefont
  {S.}~\bibnamefont {Kibbel}}, \bibinfo {author} {\bibfnamefont
  {J.}~\bibnamefont {Koch}}, \bibinfo {author} {\bibfnamefont {A.}~\bibnamefont
  {Kusnyerik}}, \bibinfo {author} {\bibfnamefont {T.}~\bibnamefont {Peters}},
  \bibinfo {author} {\bibfnamefont {K.}~\bibnamefont {Stingl}}, \bibinfo
  {author} {\bibfnamefont {H.}~\bibnamefont {Sachs}}, \bibinfo {author}
  {\bibfnamefont {A.}~\bibnamefont {Stett}}, \bibinfo {author} {\bibfnamefont
  {P.}~\bibnamefont {Szurman}}, \bibinfo {author} {\bibfnamefont
  {B.}~\bibnamefont {Wilhelm}}, \ and\ \bibinfo {author} {\bibfnamefont
  {R.}~\bibnamefont {Wilke}},\ }\bibfield  {title} {\enquote {\bibinfo {title}
  {{Subretinal electronic chips allow blind patients to read letters and
  combine them to words}},}\ }\href@noop {} {\bibfield  {journal} {\bibinfo
  {journal} {Proc. R. Soc. B}\ }\textbf {\bibinfo {volume} {278}},\ \bibinfo
  {pages} {1489} (\bibinfo {year} {2011})}\BibitemShut {NoStop}%
\bibitem [{\citenamefont {Mathieson}\ \emph {et~al.}(2012)\citenamefont
  {Mathieson}, \citenamefont {Loudin}, \citenamefont {Goetz}, \citenamefont
  {Huie}, \citenamefont {Wang}, \citenamefont {Kamins}, \citenamefont
  {Galambos}, \citenamefont {Smith}, \citenamefont {Harris}, \citenamefont
  {Sher},\ and\ \citenamefont {Palanker}}]{Mathieson12}%
  \BibitemOpen
  \bibfield  {author} {\bibinfo {author} {\bibfnamefont {K.}~\bibnamefont
  {Mathieson}}, \bibinfo {author} {\bibfnamefont {J.}~\bibnamefont {Loudin}},
  \bibinfo {author} {\bibfnamefont {G.}~\bibnamefont {Goetz}}, \bibinfo
  {author} {\bibfnamefont {P.}~\bibnamefont {Huie}}, \bibinfo {author}
  {\bibfnamefont {L.}~\bibnamefont {Wang}}, \bibinfo {author} {\bibfnamefont
  {T.~I.}\ \bibnamefont {Kamins}}, \bibinfo {author} {\bibfnamefont
  {L.}~\bibnamefont {Galambos}}, \bibinfo {author} {\bibfnamefont
  {R.}~\bibnamefont {Smith}}, \bibinfo {author} {\bibfnamefont {J.~S.}\
  \bibnamefont {Harris}}, \bibinfo {author} {\bibfnamefont {A.}~\bibnamefont
  {Sher}}, \ and\ \bibinfo {author} {\bibfnamefont {D.}~\bibnamefont
  {Palanker}},\ }\bibfield  {title} {\enquote {\bibinfo {title} {Photovoltaic
  retinal prosthesis with high pixel density},}\ }\href@noop {} {\bibfield
  {journal} {\bibinfo  {journal} {Nature Photonics}\ }\textbf {\bibinfo
  {volume} {6}},\ \bibinfo {pages} {391} (\bibinfo {year} {2012})}\BibitemShut
  {NoStop}%
\bibitem [{\citenamefont {Yue}\ \emph {et~al.}(2020)\citenamefont {Yue},
  \citenamefont {Wuyyuru}, \citenamefont {Gonzalez-Calle}, \citenamefont
  {Dorn},\ and\ \citenamefont {Humayun}}]{Yue20}%
  \BibitemOpen
  \bibfield  {author} {\bibinfo {author} {\bibfnamefont {L.}~\bibnamefont
  {Yue}}, \bibinfo {author} {\bibfnamefont {V.}~\bibnamefont {Wuyyuru}},
  \bibinfo {author} {\bibfnamefont {A.}~\bibnamefont {Gonzalez-Calle}},
  \bibinfo {author} {\bibfnamefont {J.~D.}\ \bibnamefont {Dorn}}, \ and\
  \bibinfo {author} {\bibfnamefont {M.~S.}\ \bibnamefont {Humayun}},\
  }\bibfield  {title} {\enquote {\bibinfo {title} {{Retina-electrode interface
  properties and vision restoration by two generations of retinal prostheses in
  one patient -- one in each eye}},}\ }\href@noop {} {\bibfield  {journal}
  {\bibinfo  {journal} {J. Neural Eng.}\ }\textbf {\bibinfo {volume} {17}},\
  \bibinfo {pages} {026020} (\bibinfo {year} {2020})}\BibitemShut {NoStop}%
\bibitem [{\citenamefont {Miocinovic}\ \emph {et~al.}(2013)\citenamefont
  {Miocinovic}, \citenamefont {Somayajula}, \citenamefont {Chitnis},\ and\
  \citenamefont {Vitek}}]{Miocinovic13}%
  \BibitemOpen
  \bibfield  {author} {\bibinfo {author} {\bibfnamefont {S.}~\bibnamefont
  {Miocinovic}}, \bibinfo {author} {\bibfnamefont {S.}~\bibnamefont
  {Somayajula}}, \bibinfo {author} {\bibfnamefont {S.}~\bibnamefont {Chitnis}},
  \ and\ \bibinfo {author} {\bibfnamefont {J.}~\bibnamefont {Vitek}},\
  }\bibfield  {title} {\enquote {\bibinfo {title} {{History, applications, and
  mechanisms of deep brain stimulation}},}\ }\href@noop {} {\bibfield
  {journal} {\bibinfo  {journal} {JAMA Neurol.}\ }\textbf {\bibinfo {volume}
  {70}},\ \bibinfo {pages} {163} (\bibinfo {year} {2013})}\BibitemShut
  {NoStop}%
\bibitem [{\citenamefont {\textit{et al.}}(2019)}]{Musk19}%
  \BibitemOpen
  \bibfield  {author} {\bibinfo {author} {\bibfnamefont {E.~Musk}\ \bibnamefont
  {\textit{et al.}}},\ }\bibfield  {title} {\enquote {\bibinfo {title} {An
  integrated brain-machine interface platform with thousands of channels},}\
  }\href@noop {} {\bibfield  {journal} {\bibinfo  {journal} {bioRxiv}\ ,\
  \bibinfo {pages} {703801}} (\bibinfo {year} {2019})}\BibitemShut {NoStop}%
\bibitem [{\citenamefont {Stett}\ \emph {et~al.}(2003)\citenamefont {Stett},
  \citenamefont {Egert}, \citenamefont {Guenther}, \citenamefont {Hofmann},
  \citenamefont {Meyer}, \citenamefont {Nisch},\ and\ \citenamefont
  {Haemmerle}}]{Stett03}%
  \BibitemOpen
  \bibfield  {author} {\bibinfo {author} {\bibfnamefont {A.}~\bibnamefont
  {Stett}}, \bibinfo {author} {\bibfnamefont {U.}~\bibnamefont {Egert}},
  \bibinfo {author} {\bibfnamefont {E.}~\bibnamefont {Guenther}}, \bibinfo
  {author} {\bibfnamefont {F.}~\bibnamefont {Hofmann}}, \bibinfo {author}
  {\bibfnamefont {T.}~\bibnamefont {Meyer}}, \bibinfo {author} {\bibfnamefont
  {W.}~\bibnamefont {Nisch}}, \ and\ \bibinfo {author} {\bibfnamefont
  {H.}~\bibnamefont {Haemmerle}},\ }\bibfield  {title} {\enquote {\bibinfo
  {title} {Biological application of microelectrode arrays in drug discovery
  and basic research},}\ }\href@noop {} {\bibfield  {journal} {\bibinfo
  {journal} {Anal. Bioanal. Chem.}\ }\textbf {\bibinfo {volume} {377}},\
  \bibinfo {pages} {486} (\bibinfo {year} {2003})}\BibitemShut {NoStop}%
\bibitem [{\citenamefont {Frey}\ \emph {et~al.}(2009)\citenamefont {Frey},
  \citenamefont {Egert}, \citenamefont {Heer}, \citenamefont {Hafizovic},\ and\
  \citenamefont {Hierlemann}}]{Frey09}%
  \BibitemOpen
  \bibfield  {author} {\bibinfo {author} {\bibfnamefont {U.}~\bibnamefont
  {Frey}}, \bibinfo {author} {\bibfnamefont {U.}~\bibnamefont {Egert}},
  \bibinfo {author} {\bibfnamefont {F.}~\bibnamefont {Heer}}, \bibinfo {author}
  {\bibfnamefont {S.}~\bibnamefont {Hafizovic}}, \ and\ \bibinfo {author}
  {\bibfnamefont {A.}~\bibnamefont {Hierlemann}},\ }\bibfield  {title}
  {\enquote {\bibinfo {title} {Microelectronic system for high-resolution
  mapping of extracellular electric fields applied to brain slices},}\
  }\href@noop {} {\bibfield  {journal} {\bibinfo  {journal} {Biosensors and
  Bioelectronics}\ }\textbf {\bibinfo {volume} {24}},\ \bibinfo {pages} {2191}
  (\bibinfo {year} {2009})}\BibitemShut {NoStop}%
\bibitem [{\citenamefont {Haas}\ \emph {et~al.}(2020)\citenamefont {Haas},
  \citenamefont {Rudorf}, \citenamefont {Becker}, \citenamefont {Daschner},
  \citenamefont {Drzyzga}, \citenamefont {Burkhardt},\ and\ \citenamefont
  {Stett}}]{Haas20}%
  \BibitemOpen
  \bibfield  {author} {\bibinfo {author} {\bibfnamefont {J.}~\bibnamefont
  {Haas}}, \bibinfo {author} {\bibfnamefont {R.}~\bibnamefont {Rudorf}},
  \bibinfo {author} {\bibfnamefont {M.}~\bibnamefont {Becker}}, \bibinfo
  {author} {\bibfnamefont {R.}~\bibnamefont {Daschner}}, \bibinfo {author}
  {\bibfnamefont {A.}~\bibnamefont {Drzyzga}}, \bibinfo {author} {\bibfnamefont
  {C.}~\bibnamefont {Burkhardt}}, \ and\ \bibinfo {author} {\bibfnamefont
  {A.}~\bibnamefont {Stett}},\ }\bibfield  {title} {\enquote {\bibinfo {title}
  {Sputtered iridium oxide as electrode material for subretinal stimulation},}\
  }\href@noop {} {\bibfield  {journal} {\bibinfo  {journal} {Sens. Mater.}\
  }\textbf {\bibinfo {volume} {32}},\ \bibinfo {pages} {2903} (\bibinfo {year}
  {2020})}\BibitemShut {NoStop}%
\bibitem [{\citenamefont {Maeng}\ \emph {et~al.}(2019)\citenamefont {Maeng},
  \citenamefont {Chakraborty}, \citenamefont {Geramifard}, \citenamefont
  {Kang}, \citenamefont {Rihani}, \citenamefont {Joshi-Imre},\ and\
  \citenamefont {Cogan}}]{Maeng19}%
  \BibitemOpen
  \bibfield  {author} {\bibinfo {author} {\bibfnamefont {J.}~\bibnamefont
  {Maeng}}, \bibinfo {author} {\bibfnamefont {B.}~\bibnamefont {Chakraborty}},
  \bibinfo {author} {\bibfnamefont {N.}~\bibnamefont {Geramifard}}, \bibinfo
  {author} {\bibfnamefont {T.}~\bibnamefont {Kang}}, \bibinfo {author}
  {\bibfnamefont {R.~T.}\ \bibnamefont {Rihani}}, \bibinfo {author}
  {\bibfnamefont {A.}~\bibnamefont {Joshi-Imre}}, \ and\ \bibinfo {author}
  {\bibfnamefont {S.~F.}\ \bibnamefont {Cogan}},\ }\bibfield  {title} {\enquote
  {\bibinfo {title} {{High-charge-capacity sputtered iridium oxide neural
  stimulation electrodes deposited using water vapor as a reactive plasma
  constituent}},}\ }\href@noop {} {\bibfield  {journal} {\bibinfo  {journal}
  {J. Biomed. Mater. Res.}\ }\textbf {\bibinfo {volume} {108B}},\ \bibinfo
  {pages} {880} (\bibinfo {year} {2019})}\BibitemShut {NoStop}%
\bibitem [{\citenamefont {Merrill}\ \emph {et~al.}(2005)\citenamefont
  {Merrill}, \citenamefont {Bikson},\ and\ \citenamefont
  {Jefferys}}]{Merrill05}%
  \BibitemOpen
  \bibfield  {author} {\bibinfo {author} {\bibfnamefont {D.~R.}\ \bibnamefont
  {Merrill}}, \bibinfo {author} {\bibfnamefont {M.}~\bibnamefont {Bikson}}, \
  and\ \bibinfo {author} {\bibfnamefont {J.~G.~R.}\ \bibnamefont {Jefferys}},\
  }\bibfield  {title} {\enquote {\bibinfo {title} {Electrical stimulation of
  excitable tissue: design of efficacious and safe protocols},}\ }\href@noop {}
  {\bibfield  {journal} {\bibinfo  {journal} {J. Neurosci. Methods}\ }\textbf
  {\bibinfo {volume} {141}},\ \bibinfo {pages} {171} (\bibinfo {year}
  {2005})}\BibitemShut {NoStop}%
\bibitem [{\citenamefont {Fromherz}\ and\ \citenamefont
  {Stett}(1995)}]{Fromherz1995}%
  \BibitemOpen
  \bibfield  {author} {\bibinfo {author} {\bibfnamefont {P.}~\bibnamefont
  {Fromherz}}\ and\ \bibinfo {author} {\bibfnamefont {A.}~\bibnamefont
  {Stett}},\ }\bibfield  {title} {\enquote {\bibinfo {title} {{Silicon-Neuron
  Junction: Capacitive Stimulation of an Individual Neuron on a Silicon
  Chip}},}\ }\href@noop {} {\bibfield  {journal} {\bibinfo  {journal} {Phys.
  Rev. Lett.}\ }\textbf {\bibinfo {volume} {75}},\ \bibinfo {pages} {1670}
  (\bibinfo {year} {1995})}\BibitemShut {NoStop}%
\bibitem [{\citenamefont {Fromherz}\ \emph {et~al.}(1991)\citenamefont
  {Fromherz}, \citenamefont {Offenh{\"a}usser}, \citenamefont {Vetter},\ and\
  \citenamefont {Weis}}]{Fromherz1991}%
  \BibitemOpen
  \bibfield  {author} {\bibinfo {author} {\bibfnamefont {P.}~\bibnamefont
  {Fromherz}}, \bibinfo {author} {\bibfnamefont {A.}~\bibnamefont
  {Offenh{\"a}usser}}, \bibinfo {author} {\bibfnamefont {T.}~\bibnamefont
  {Vetter}}, \ and\ \bibinfo {author} {\bibfnamefont {J.}~\bibnamefont
  {Weis}},\ }\bibfield  {title} {\enquote {\bibinfo {title} {{A Neuron-Silicon
  Junction: A Retzius Cell of the Leech on an Insulated-Gate Field-Effect
  Transistor}},}\ }\href@noop {} {\bibfield  {journal} {\bibinfo  {journal}
  {Science}\ }\textbf {\bibinfo {volume} {252}},\ \bibinfo {pages} {1290}
  (\bibinfo {year} {1991})}\BibitemShut {NoStop}%
\bibitem [{\citenamefont {Bertotti}\ \emph {et~al.}(2014)\citenamefont
  {Bertotti}, \citenamefont {Velychko}, \citenamefont {Dodel}, \citenamefont
  {Keil}, \citenamefont {Wolansky}, \citenamefont {Tillak}, \citenamefont
  {Schreiter}, \citenamefont {Grall}, \citenamefont {Jesinger}, \citenamefont
  {R{\"o}hler}, \citenamefont {Eickenscheidt}, \citenamefont {Stett},
  \citenamefont {M{\"o}ller}, \citenamefont {Boven}, \citenamefont {Zeck},\
  and\ \citenamefont {Thewes}}]{Bertotti14}%
  \BibitemOpen
  \bibfield  {author} {\bibinfo {author} {\bibfnamefont {G.}~\bibnamefont
  {Bertotti}}, \bibinfo {author} {\bibfnamefont {D.}~\bibnamefont {Velychko}},
  \bibinfo {author} {\bibfnamefont {N.}~\bibnamefont {Dodel}}, \bibinfo
  {author} {\bibfnamefont {S.}~\bibnamefont {Keil}}, \bibinfo {author}
  {\bibfnamefont {D.}~\bibnamefont {Wolansky}}, \bibinfo {author}
  {\bibfnamefont {B.}~\bibnamefont {Tillak}}, \bibinfo {author} {\bibfnamefont
  {M.}~\bibnamefont {Schreiter}}, \bibinfo {author} {\bibfnamefont
  {A.}~\bibnamefont {Grall}}, \bibinfo {author} {\bibfnamefont
  {P.}~\bibnamefont {Jesinger}}, \bibinfo {author} {\bibfnamefont
  {S.}~\bibnamefont {R{\"o}hler}}, \bibinfo {author} {\bibfnamefont
  {M.}~\bibnamefont {Eickenscheidt}}, \bibinfo {author} {\bibfnamefont
  {A.}~\bibnamefont {Stett}}, \bibinfo {author} {\bibfnamefont
  {A.}~\bibnamefont {M{\"o}ller}}, \bibinfo {author} {\bibfnamefont {K-H.}\
  \bibnamefont {Boven}}, \bibinfo {author} {\bibfnamefont {G.}~\bibnamefont
  {Zeck}}, \ and\ \bibinfo {author} {\bibfnamefont {R.}~\bibnamefont
  {Thewes}},\ }\bibfield  {title} {\enquote {\bibinfo {title} {{A CMOS-Based
  Sensor Array for In-Vitro Neural Tissue Interfacing with 4225 Recording Sites
  and 1024 Stimulation Sites}},}\ }in\ \href@noop {} {\emph {\bibinfo
  {booktitle} {2014 IEEE Biomedical Circuits and Systems Conference (BioCAS)
  Proceedings}}}\ (\bibinfo {year} {2014})\BibitemShut {NoStop}%
\bibitem [{\citenamefont {Wallrapp}\ and\ \citenamefont
  {Fromherz}(2006)}]{Wallrapp06}%
  \BibitemOpen
  \bibfield  {author} {\bibinfo {author} {\bibfnamefont {F.}~\bibnamefont
  {Wallrapp}}\ and\ \bibinfo {author} {\bibfnamefont {P.}~\bibnamefont
  {Fromherz}},\ }\bibfield  {title} {\enquote {\bibinfo {title} {{TiO$_2$} and
  {HfO$_2$} in electrolyte-oxide-silicon configuration for applications in
  bioelectronics},}\ }\href@noop {} {\bibfield  {journal} {\bibinfo  {journal}
  {J. Appl. Phys.}\ }\textbf {\bibinfo {volume} {99}},\ \bibinfo {pages}
  {114103} (\bibinfo {year} {2006})}\BibitemShut {NoStop}%
\bibitem [{\citenamefont {Schoen}\ and\ \citenamefont
  {Fromherz}(2007)}]{Schoen07}%
  \BibitemOpen
  \bibfield  {author} {\bibinfo {author} {\bibfnamefont {I.}~\bibnamefont
  {Schoen}}\ and\ \bibinfo {author} {\bibfnamefont {P.}~\bibnamefont
  {Fromherz}},\ }\bibfield  {title} {\enquote {\bibinfo {title} {The mechanism
  of extracellular stimulation of nerve cells on an
  electrolyte-oxide-semiconductor capacitor},}\ }\href@noop {} {\bibfield
  {journal} {\bibinfo  {journal} {Biophysical Journal}\ }\textbf {\bibinfo
  {volume} {92}},\ \bibinfo {pages} {1096} (\bibinfo {year}
  {2007})}\BibitemShut {NoStop}%
\bibitem [{\citenamefont {Eickenscheidt}\ \emph {et~al.}(2012)\citenamefont
  {Eickenscheidt}, \citenamefont {Jenkner}, \citenamefont {Thewes},
  \citenamefont {Fromherz},\ and\ \citenamefont {Zeck}}]{Eickenscheidt12}%
  \BibitemOpen
  \bibfield  {author} {\bibinfo {author} {\bibfnamefont {M.}~\bibnamefont
  {Eickenscheidt}}, \bibinfo {author} {\bibfnamefont {M.}~\bibnamefont
  {Jenkner}}, \bibinfo {author} {\bibfnamefont {R.}~\bibnamefont {Thewes}},
  \bibinfo {author} {\bibfnamefont {P.}~\bibnamefont {Fromherz}}, \ and\
  \bibinfo {author} {\bibfnamefont {G.}~\bibnamefont {Zeck}},\ }\bibfield
  {title} {\enquote {\bibinfo {title} {Electrical stimulation of retinal
  neurons in epiretinal and subretinal configuration using a multicapacitor
  array},}\ }\href@noop {} {\bibfield  {journal} {\bibinfo  {journal} {J.
  Neurophysiol.}\ }\textbf {\bibinfo {volume} {107}},\ \bibinfo {pages} {2742}
  (\bibinfo {year} {2012})}\BibitemShut {NoStop}%
\bibitem [{\citenamefont {Dollt}\ \emph {et~al.}(2020)\citenamefont {Dollt},
  \citenamefont {Reh}, \citenamefont {Metzger}, \citenamefont {Heusel},
  \citenamefont {Kriebel}, \citenamefont {Bucher},\ and\ \citenamefont
  {Zeck}}]{Dollt20}%
  \BibitemOpen
  \bibfield  {author} {\bibinfo {author} {\bibfnamefont {M.}~\bibnamefont
  {Dollt}}, \bibinfo {author} {\bibfnamefont {M.}~\bibnamefont {Reh}}, \bibinfo
  {author} {\bibfnamefont {M.}~\bibnamefont {Metzger}}, \bibinfo {author}
  {\bibfnamefont {G.}~\bibnamefont {Heusel}}, \bibinfo {author} {\bibfnamefont
  {M.}~\bibnamefont {Kriebel}}, \bibinfo {author} {\bibfnamefont
  {V.}~\bibnamefont {Bucher}}, \ and\ \bibinfo {author} {\bibfnamefont
  {G.}~\bibnamefont {Zeck}},\ }\bibfield  {title} {\enquote {\bibinfo {title}
  {{Low-Temperature Atomic Layer Deposited Oxide on Titanium Nitride Electrodes
  Enables Culture and Physiological Recording of Electrogenic Cells}},}\
  }\href@noop {} {\bibfield  {journal} {\bibinfo  {journal} {Front. Neurosci.}\
  }\textbf {\bibinfo {volume} {14}},\ \bibinfo {pages} {552876} (\bibinfo
  {year} {2020})}\BibitemShut {NoStop}%
\bibitem [{\citenamefont {Corna}\ \emph {et~al.}(2018)\citenamefont {Corna},
  \citenamefont {Herrmann},\ and\ \citenamefont {Zeck}}]{Corna18}%
  \BibitemOpen
  \bibfield  {author} {\bibinfo {author} {\bibfnamefont {A.}~\bibnamefont
  {Corna}}, \bibinfo {author} {\bibfnamefont {T.}~\bibnamefont {Herrmann}}, \
  and\ \bibinfo {author} {\bibfnamefont {G.}~\bibnamefont {Zeck}},\ }\bibfield
  {title} {\enquote {\bibinfo {title} {Electrode-size dependent thresholds in
  subretinal neuroprosthetic stimulation},}\ }\href@noop {} {\bibfield
  {journal} {\bibinfo  {journal} {J. Neural Eng.}\ }\textbf {\bibinfo {volume}
  {15}},\ \bibinfo {pages} {045003} (\bibinfo {year} {2018})}\BibitemShut
  {NoStop}%
\bibitem [{\citenamefont {Brandt}\ and\ \citenamefont
  {Dahmen}(2005)}]{Brandt_Book05}%
  \BibitemOpen
  \bibfield  {author} {\bibinfo {author} {\bibfnamefont {S.}~\bibnamefont
  {Brandt}}\ and\ \bibinfo {author} {\bibfnamefont {H.~D.}\ \bibnamefont
  {Dahmen}},\ }\href@noop {} {\emph {\bibinfo {title} {Elektrodynamik}}},\
  \bibinfo {edition} {4th}\ ed.\ (\bibinfo  {publisher} {Springer},\ \bibinfo
  {address} {Berlin},\ \bibinfo {year} {2005})\BibitemShut {NoStop}%
\bibitem [{\citenamefont {Waser}\ \emph {et~al.}(2005)\citenamefont {Waser},
  \citenamefont {B{\"o}ttger},\ and\ \citenamefont {(Ed.)}}]{Waser_Book05}%
  \BibitemOpen
  \bibfield  {author} {\bibinfo {author} {\bibfnamefont {R.}~\bibnamefont
  {Waser}}, \bibinfo {author} {\bibfnamefont {U.}~\bibnamefont {B{\"o}ttger}},
  \ and\ \bibinfo {author} {\bibfnamefont {S.~Tiedke}\ \bibnamefont {(Ed.)}},\
  }\href@noop {} {\emph {\bibinfo {title} {Polar Oxides}}},\ \bibinfo {edition}
  {1st}\ ed.\ (\bibinfo  {publisher} {Wiley},\ \bibinfo {address} {New York},\
  \bibinfo {year} {2005})\BibitemShut {NoStop}%
\bibitem [{\citenamefont {Park}\ \emph {et~al.}(2018)\citenamefont {Park},
  \citenamefont {Lee}, \citenamefont {Mikolajick}, \citenamefont {Schroeder},\
  and\ \citenamefont {Hwang}}]{Park18}%
  \BibitemOpen
  \bibfield  {author} {\bibinfo {author} {\bibfnamefont {M.~H.}\ \bibnamefont
  {Park}}, \bibinfo {author} {\bibfnamefont {Y.~H.}\ \bibnamefont {Lee}},
  \bibinfo {author} {\bibfnamefont {T.}~\bibnamefont {Mikolajick}}, \bibinfo
  {author} {\bibfnamefont {U.}~\bibnamefont {Schroeder}}, \ and\ \bibinfo
  {author} {\bibfnamefont {C.~S.}\ \bibnamefont {Hwang}},\ }\bibfield  {title}
  {\enquote {\bibinfo {title} {{Review and perspective on ferroelectric
  HfO$_2$-based thin films for memory applications}},}\ }\href@noop {}
  {\bibfield  {journal} {\bibinfo  {journal} {MRS Communications}\ }\textbf
  {\bibinfo {volume} {8}},\ \bibinfo {pages} {795} (\bibinfo {year}
  {2018})}\BibitemShut {NoStop}%
\bibitem [{\citenamefont {M{\"u}ller}\ \emph {et~al.}(2011)\citenamefont
  {M{\"u}ller}, \citenamefont {B{\"o}schke}, \citenamefont {Br{\"a}uhaus},
  \citenamefont {Schr{\"o}der}, \citenamefont {B{\"o}ttger}, \citenamefont
  {Sundqvist}, \citenamefont {K{\"u}cher}, \citenamefont {Mikolajick},\ and\
  \citenamefont {Frey}}]{Mueller11}%
  \BibitemOpen
  \bibfield  {author} {\bibinfo {author} {\bibfnamefont {J.}~\bibnamefont
  {M{\"u}ller}}, \bibinfo {author} {\bibfnamefont {T.~S.}\ \bibnamefont
  {B{\"o}schke}}, \bibinfo {author} {\bibfnamefont {D.}~\bibnamefont
  {Br{\"a}uhaus}}, \bibinfo {author} {\bibfnamefont {U.}~\bibnamefont
  {Schr{\"o}der}}, \bibinfo {author} {\bibfnamefont {U.}~\bibnamefont
  {B{\"o}ttger}}, \bibinfo {author} {\bibfnamefont {J.}~\bibnamefont
  {Sundqvist}}, \bibinfo {author} {\bibfnamefont {P.}~\bibnamefont
  {K{\"u}cher}}, \bibinfo {author} {\bibfnamefont {T.}~\bibnamefont
  {Mikolajick}}, \ and\ \bibinfo {author} {\bibfnamefont {L.}~\bibnamefont
  {Frey}},\ }\bibfield  {title} {\enquote {\bibinfo {title} {Ferroelectric
  {Zr$_{0.5}$Hf$_{0.5}$O$_{2}$} thin films for nonvolatile memory
  applications},}\ }\href@noop {} {\bibfield  {journal} {\bibinfo  {journal}
  {Appl. Phys. Lett.}\ }\textbf {\bibinfo {volume} {99}},\ \bibinfo {pages}
  {112901} (\bibinfo {year} {2011})}\BibitemShut {NoStop}%
\bibitem [{\citenamefont {Wei}\ \emph {et~al.}(2018)\citenamefont {Wei},
  \citenamefont {Nukala}, \citenamefont {Salverda}, \citenamefont {Matzen},
  \citenamefont {Zhao}, \citenamefont {Momand}, \citenamefont {Everhardt},
  \citenamefont {Agnus}, \citenamefont {Blake}, \citenamefont {Lecoeur},
  \citenamefont {Kooi}, \citenamefont {{\'{I}}{\~{n}}iguez}, \citenamefont
  {Dkhil},\ and\ \citenamefont {Noheda}}]{Wei18}%
  \BibitemOpen
  \bibfield  {author} {\bibinfo {author} {\bibfnamefont {Y.}~\bibnamefont
  {Wei}}, \bibinfo {author} {\bibfnamefont {P.}~\bibnamefont {Nukala}},
  \bibinfo {author} {\bibfnamefont {M.}~\bibnamefont {Salverda}}, \bibinfo
  {author} {\bibfnamefont {S.}~\bibnamefont {Matzen}}, \bibinfo {author}
  {\bibfnamefont {H.~J.}\ \bibnamefont {Zhao}}, \bibinfo {author}
  {\bibfnamefont {J.}~\bibnamefont {Momand}}, \bibinfo {author} {\bibfnamefont
  {A.~S.}\ \bibnamefont {Everhardt}}, \bibinfo {author} {\bibfnamefont
  {G.}~\bibnamefont {Agnus}}, \bibinfo {author} {\bibfnamefont {G.~R.}\
  \bibnamefont {Blake}}, \bibinfo {author} {\bibfnamefont {P.}~\bibnamefont
  {Lecoeur}}, \bibinfo {author} {\bibfnamefont {B.~J.}\ \bibnamefont {Kooi}},
  \bibinfo {author} {\bibfnamefont {J.}~\bibnamefont {{\'{I}}{\~{n}}iguez}},
  \bibinfo {author} {\bibfnamefont {B.}~\bibnamefont {Dkhil}}, \ and\ \bibinfo
  {author} {\bibfnamefont {B.}~\bibnamefont {Noheda}},\ }\bibfield  {title}
  {\enquote {\bibinfo {title} {{A rhombohedral ferroelectric phase in
  epitaxially strained Hf$_{0.5}$Zr$_{0.5}$O$_2$ thin films}},}\ }\href@noop {}
  {\bibfield  {journal} {\bibinfo  {journal} {Nature Materials}\ }\textbf
  {\bibinfo {volume} {17}},\ \bibinfo {pages} {1095} (\bibinfo {year}
  {2018})}\BibitemShut {NoStop}%
\bibitem [{\citenamefont {Damjanovic}(1998)}]{Damjanovic1998}%
  \BibitemOpen
  \bibfield  {author} {\bibinfo {author} {\bibfnamefont {D.}~\bibnamefont
  {Damjanovic}},\ }\bibfield  {title} {\enquote {\bibinfo {title}
  {Ferroelectric, dielectric and piezoelectric properties of ferroelectric thin
  films and ceramics},}\ }\href@noop {} {\bibfield  {journal} {\bibinfo
  {journal} {Rep. Progr. Phys.}\ }\textbf {\bibinfo {volume} {61}},\ \bibinfo
  {pages} {1267} (\bibinfo {year} {1998})}\BibitemShut {NoStop}%
\bibitem [{\citenamefont {Miga}\ \emph {et~al.}(2007)\citenamefont {Miga},
  \citenamefont {Dec},\ and\ \citenamefont {Kleemann}}]{Miga07}%
  \BibitemOpen
  \bibfield  {author} {\bibinfo {author} {\bibfnamefont {S.}~\bibnamefont
  {Miga}}, \bibinfo {author} {\bibfnamefont {J.}~\bibnamefont {Dec}}, \ and\
  \bibinfo {author} {\bibfnamefont {W.}~\bibnamefont {Kleemann}},\ }\bibfield
  {title} {\enquote {\bibinfo {title} {{Computer-controlled susceptometer for
  investigating the linear and nonlinear dielectric response}},}\ }\href@noop
  {} {\bibfield  {journal} {\bibinfo  {journal} {Rev. Sci. Instrum.}\ }\textbf
  {\bibinfo {volume} {78}},\ \bibinfo {pages} {033902} (\bibinfo {year}
  {2007})}\BibitemShut {NoStop}%
\bibitem [{\citenamefont {Freeman}\ \emph {et~al.}(2010)\citenamefont
  {Freeman}, \citenamefont {Eddington}, \citenamefont {III},\ and\
  \citenamefont {Fried}}]{Freeman10}%
  \BibitemOpen
  \bibfield  {author} {\bibinfo {author} {\bibfnamefont {D.~K.}\ \bibnamefont
  {Freeman}}, \bibinfo {author} {\bibfnamefont {D.~K.}\ \bibnamefont
  {Eddington}}, \bibinfo {author} {\bibfnamefont {J.~F.~Rizzo}\ \bibnamefont
  {III}}, \ and\ \bibinfo {author} {\bibfnamefont {S.~I.}\ \bibnamefont
  {Fried}},\ }\bibfield  {title} {\enquote {\bibinfo {title} {{Selective
  activation of neuronal targets with sinusoidal electric stimulation}},}\
  }\href@noop {} {\bibfield  {journal} {\bibinfo  {journal} {J. Neurophysiol.}\
  }\textbf {\bibinfo {volume} {104}},\ \bibinfo {pages} {2778} (\bibinfo {year}
  {2010})}\BibitemShut {NoStop}%
\bibitem [{\citenamefont {Becker}\ \emph {et~al.}(2020)\citenamefont {Becker},
  \citenamefont {Burkhardt}, \citenamefont {Schr{\"o}ppel}, \citenamefont
  {Kleiner},\ and\ \citenamefont {Koelle}}]{BeckerJAP20}%
  \BibitemOpen
  \bibfield  {author} {\bibinfo {author} {\bibfnamefont {M.}~\bibnamefont
  {Becker}}, \bibinfo {author} {\bibfnamefont {C.~J.}\ \bibnamefont
  {Burkhardt}}, \bibinfo {author} {\bibfnamefont {B.}~\bibnamefont
  {Schr{\"o}ppel}}, \bibinfo {author} {\bibfnamefont {R.}~\bibnamefont
  {Kleiner}}, \ and\ \bibinfo {author} {\bibfnamefont {D.}~\bibnamefont
  {Koelle}},\ }\bibfield  {title} {\enquote {\bibinfo {title} {Rayleigh
  analysis and dielectric dispersion in polycrystalline
  {0.5(Ba$_{0.7}$Ca$_{0.3}$)TiO$_{3}$--0.5Ba(Zr$_{0.2}$Ti$_{0.8}$)O$_{3}$}
  ferroelectric thin films by domain-wall pinning element modeling},}\
  }\href@noop {} {\bibfield  {journal} {\bibinfo  {journal} {J. Appl. Phys.}\
  }\textbf {\bibinfo {volume} {128}},\ \bibinfo {pages} {154103} (\bibinfo
  {year} {2020})}\BibitemShut {NoStop}%
\bibitem [{\citenamefont {Jo}\ \emph {et~al.}(2007)\citenamefont {Jo},
  \citenamefont {Han}, \citenamefont {Yoon}, \citenamefont {Song},
  \citenamefont {Kim},\ and\ \citenamefont {Noh}}]{Jo07}%
  \BibitemOpen
  \bibfield  {author} {\bibinfo {author} {\bibfnamefont {J.~Y.}\ \bibnamefont
  {Jo}}, \bibinfo {author} {\bibfnamefont {H.~S.}\ \bibnamefont {Han}},
  \bibinfo {author} {\bibfnamefont {J.-G.}\ \bibnamefont {Yoon}}, \bibinfo
  {author} {\bibfnamefont {T.~K.}\ \bibnamefont {Song}}, \bibinfo {author}
  {\bibfnamefont {S.-H.}\ \bibnamefont {Kim}}, \ and\ \bibinfo {author}
  {\bibfnamefont {T.~W.}\ \bibnamefont {Noh}},\ }\bibfield  {title} {\enquote
  {\bibinfo {title} {{Domain switching kinetics in disordered ferroelectric
  thin films}},}\ }\href@noop {} {\bibfield  {journal} {\bibinfo  {journal}
  {Phys. Rev. Lett.}\ }\textbf {\bibinfo {volume} {99}},\ \bibinfo {pages}
  {267602} (\bibinfo {year} {2007})}\BibitemShut {NoStop}%
\bibitem [{\citenamefont {Kolmogorov}(1937)}]{Kolmogorov1937}%
  \BibitemOpen
  \bibfield  {author} {\bibinfo {author} {\bibfnamefont {A.~N.}\ \bibnamefont
  {Kolmogorov}},\ }\bibfield  {title} {\enquote {\bibinfo {title} {{Statistical
  theory of nucleation processes}},}\ }\href@noop {} {\bibfield  {journal}
  {\bibinfo  {journal} {Bull. Acad. Sci. USSR Math. Ser.}\ }\textbf {\bibinfo
  {volume} {3}},\ \bibinfo {pages} {355} (\bibinfo {year} {1937})}\BibitemShut
  {NoStop}%
\bibitem [{\citenamefont {Avrami}(1939)}]{Avrami1939}%
  \BibitemOpen
  \bibfield  {author} {\bibinfo {author} {\bibfnamefont {M.}~\bibnamefont
  {Avrami}},\ }\bibfield  {title} {\enquote {\bibinfo {title} {{Kinetics of
  phase change. I General theory}},}\ }\href@noop {} {\bibfield  {journal}
  {\bibinfo  {journal} {J. Chem. Phys.}\ }\textbf {\bibinfo {volume} {7}},\
  \bibinfo {pages} {1103} (\bibinfo {year} {1939})}\BibitemShut {NoStop}%
\bibitem [{\citenamefont {Ishibashi}\ and\ \citenamefont
  {Takagi}(1971)}]{Ishibashi1971}%
  \BibitemOpen
  \bibfield  {author} {\bibinfo {author} {\bibfnamefont {Y.}~\bibnamefont
  {Ishibashi}}\ and\ \bibinfo {author} {\bibfnamefont {Y.}~\bibnamefont
  {Takagi}},\ }\bibfield  {title} {\enquote {\bibinfo {title} {{Note on
  ferroelectric domain switching}},}\ }\href@noop {} {\bibfield  {journal}
  {\bibinfo  {journal} {J. Phys. Soc. Jpn.}\ }\textbf {\bibinfo {volume}
  {31}},\ \bibinfo {pages} {506} (\bibinfo {year} {1971})}\BibitemShut
  {NoStop}%
\bibitem [{\citenamefont {Chen}\ \emph {et~al.}(2020)\citenamefont {Chen},
  \citenamefont {Ryzhik},\ and\ \citenamefont {Palanker}}]{Chen20}%
  \BibitemOpen
  \bibfield  {author} {\bibinfo {author} {\bibfnamefont {Z.}~\bibnamefont
  {Chen}}, \bibinfo {author} {\bibfnamefont {L.}~\bibnamefont {Ryzhik}}, \ and\
  \bibinfo {author} {\bibfnamefont {D.}~\bibnamefont {Palanker}},\ }\bibfield
  {title} {\enquote {\bibinfo {title} {Current distribution on capacitive
  electrode-electrolyte interfaces},}\ }\href@noop {} {\bibfield  {journal}
  {\bibinfo  {journal} {Phys. Rev. Applied}\ }\textbf {\bibinfo {volume}
  {13}},\ \bibinfo {pages} {014004} (\bibinfo {year} {2020})}\BibitemShut
  {NoStop}%
\bibitem [{\citenamefont {Ferris}\ \emph {et~al.}(2013)\citenamefont {Ferris},
  \citenamefont {Lin}, \citenamefont {Therezien}, \citenamefont {Yellen},\ and\
  \citenamefont {Zauscher}}]{Ferris13}%
  \BibitemOpen
  \bibfield  {author} {\bibinfo {author} {\bibfnamefont {R.~J.}\ \bibnamefont
  {Ferris}}, \bibinfo {author} {\bibfnamefont {S.}~\bibnamefont {Lin}},
  \bibinfo {author} {\bibfnamefont {M.}~\bibnamefont {Therezien}}, \bibinfo
  {author} {\bibfnamefont {B.~B.}\ \bibnamefont {Yellen}}, \ and\ \bibinfo
  {author} {\bibfnamefont {S.}~\bibnamefont {Zauscher}},\ }\bibfield  {title}
  {\enquote {\bibinfo {title} {Electric double layer formed by polarized
  ferroelectric thin films},}\ }\href@noop {} {\bibfield  {journal} {\bibinfo
  {journal} {ACS Appl. Mater. Interfaces}\ }\textbf {\bibinfo {volume} {5}},\
  \bibinfo {pages} {2610} (\bibinfo {year} {2013})}\BibitemShut {NoStop}%
\bibitem [{\citenamefont {Fabiano}\ \emph {et~al.}(2014)\citenamefont
  {Fabiano}, \citenamefont {Crispin},\ and\ \citenamefont
  {Berggren}}]{Fabiano14}%
  \BibitemOpen
  \bibfield  {author} {\bibinfo {author} {\bibfnamefont {S.}~\bibnamefont
  {Fabiano}}, \bibinfo {author} {\bibfnamefont {X.}~\bibnamefont {Crispin}}, \
  and\ \bibinfo {author} {\bibfnamefont {M.}~\bibnamefont {Berggren}},\
  }\bibfield  {title} {\enquote {\bibinfo {title} {{Ferroelectric Polarization
  Induces Electric Double Layer Bistability in Electrolyte-Gated Field-Effect
  Transistors}},}\ }\href@noop {} {\bibfield  {journal} {\bibinfo  {journal}
  {ACS Appl. Mater. Interfaces}\ }\textbf {\bibinfo {volume} {6}},\ \bibinfo
  {pages} {438} (\bibinfo {year} {2014})}\BibitemShut {NoStop}%
\bibitem [{\citenamefont {Toss}\ \emph {et~al.}(2017)\citenamefont {Toss},
  \citenamefont {L{\"o}nnqvist}, \citenamefont {Nilsson}, \citenamefont
  {Sawatdee}, \citenamefont {Nissa}, \citenamefont {Fabiano}, \citenamefont
  {Berggren}, \citenamefont {Kratz},\ and\ \citenamefont {Simon}}]{Toss17}%
  \BibitemOpen
  \bibfield  {author} {\bibinfo {author} {\bibfnamefont {H.}~\bibnamefont
  {Toss}}, \bibinfo {author} {\bibfnamefont {S.}~\bibnamefont {L{\"o}nnqvist}},
  \bibinfo {author} {\bibfnamefont {D.}~\bibnamefont {Nilsson}}, \bibinfo
  {author} {\bibfnamefont {A.}~\bibnamefont {Sawatdee}}, \bibinfo {author}
  {\bibfnamefont {J.}~\bibnamefont {Nissa}}, \bibinfo {author} {\bibfnamefont
  {S.}~\bibnamefont {Fabiano}}, \bibinfo {author} {\bibfnamefont
  {M.}~\bibnamefont {Berggren}}, \bibinfo {author} {\bibfnamefont
  {G.}~\bibnamefont {Kratz}}, \ and\ \bibinfo {author} {\bibfnamefont {D.~T.}\
  \bibnamefont {Simon}},\ }\bibfield  {title} {\enquote {\bibinfo {title}
  {Ferroelectric surfaces for cell release},}\ }\href@noop {} {\bibfield
  {journal} {\bibinfo  {journal} {Synthetic Metals}\ }\textbf {\bibinfo
  {volume} {228}},\ \bibinfo {pages} {99} (\bibinfo {year} {2017})}\BibitemShut
  {NoStop}%
\bibitem [{\citenamefont {Fichtner}\ \emph {et~al.}(2019)\citenamefont
  {Fichtner}, \citenamefont {Wolff}, \citenamefont {Lofink}, \citenamefont
  {Kienzle},\ and\ \citenamefont {Wagner}}]{Fichtner19}%
  \BibitemOpen
  \bibfield  {author} {\bibinfo {author} {\bibfnamefont {S.}~\bibnamefont
  {Fichtner}}, \bibinfo {author} {\bibfnamefont {N.}~\bibnamefont {Wolff}},
  \bibinfo {author} {\bibfnamefont {F.}~\bibnamefont {Lofink}}, \bibinfo
  {author} {\bibfnamefont {L.}~\bibnamefont {Kienzle}}, \ and\ \bibinfo
  {author} {\bibfnamefont {B.}~\bibnamefont {Wagner}},\ }\bibfield  {title}
  {\enquote {\bibinfo {title} {{AlScN: A III-V semiconductor based
  ferroelectric}},}\ }\href@noop {} {\bibfield  {journal} {\bibinfo  {journal}
  {J. Appl. Phys.}\ }\textbf {\bibinfo {volume} {125}},\ \bibinfo {pages}
  {114103} (\bibinfo {year} {2019})}\BibitemShut {NoStop}%
\bibitem [{\citenamefont {Mulaosmanovic}\ \emph {et~al.}(2018)\citenamefont
  {Mulaosmanovic}, \citenamefont {Chicca}, \citenamefont {Bertele},
  \citenamefont {Mikolajick},\ and\ \citenamefont
  {Slesazeck}}]{Mulaosmanovic18}%
  \BibitemOpen
  \bibfield  {author} {\bibinfo {author} {\bibfnamefont {H.}~\bibnamefont
  {Mulaosmanovic}}, \bibinfo {author} {\bibfnamefont {E.}~\bibnamefont
  {Chicca}}, \bibinfo {author} {\bibfnamefont {M.}~\bibnamefont {Bertele}},
  \bibinfo {author} {\bibfnamefont {T.}~\bibnamefont {Mikolajick}}, \ and\
  \bibinfo {author} {\bibfnamefont {S.}~\bibnamefont {Slesazeck}},\ }\bibfield
  {title} {\enquote {\bibinfo {title} {Mimicking biological neurons with a
  nanoscale ferroelectric transistor},}\ }\href@noop {} {\bibfield  {journal}
  {\bibinfo  {journal} {Nanoscale}\ }\textbf {\bibinfo {volume} {10}},\
  \bibinfo {pages} {21755} (\bibinfo {year} {2018})}\BibitemShut {NoStop}%
\end{thebibliography}%
\end{document}